\documentclass[twocolumn,12pt,a4paper]{iopart}

\usepackage{graphicx}
\usepackage{dcolumn}
\usepackage{bm}
\usepackage{color}
\usepackage{multirow}
\usepackage{amssymb}
\usepackage{cite}

\begin{document}
 
\title{Scaling in the correlation energies of two-dimensional artificial atoms}

\author{
Alexander Odriazola$^{1,2,3}$,\ Mikko M Ervasti$^{1,3}$,\ Ilja Makkonen$^{1,3}$,\ Alain Delgado$^{5,6}$,\ Augusto Gonz\'{a}lez$^{4}$, Esa R\"as\"anen$^{2}$, and Ari Harju$^{1,3}$
}

\address{
$^{1}$COMP Centre of Excellence, Department of Applied Physics, Aalto University School of Science, PO Box 14100, FI-00076 AALTO, Espoo, Finland}
\address{$^{2}$Department of Physics, Tampere University of Technology, FI-33101 Tampere, Finland}
\address{$^{3}$Helsinki Institute of Physics, Aalto University, PO Box 14100, FI-00076 AALTO, Espoo, Finland}
\address{$^{4}$Institute of Cybernetics Mathematics and Physics (ICIMAF), Calle E \#309, CP 10400, Havana, Cuba}
\address{$^{5}$CNR-NANO S3, Institute for Nanoscience, Via Campi 213/A 41125, Modena, Italy}
\address{$^{6}$Centro de Aplicaciones Tecnol\'{o}gicas y Desarrollo Nuclear (CEADEN), Calle 30 \#502, CP 11300, Havana, Cuba}

\begin{abstract}
We find an unexpected scaling in the correlation energy of artificial atoms, i.e., 
harmonically confined two-dimensional quantum dots. The scaling relation is found
through extensive numerical examinations including Hartree-Fock, variational quantum 
Monte Carlo, density-functional, and full configuration-interaction calculations.
We show that the correlation energy, i.e., the true ground-state total energy subtracted by
the Hartree-Fock total energy, follows a simple function of the Coulomb energy,
confimenent strength and, the number of electrons. We find an analytic expression for
this function, as well as for the correlation energy per particle and for the 
ratio between the correlation and total energies. Our tests for independent
diffusion Monte Carlo and coupled-cluster results for quantum dots -- including
open-shell data -- confirm the generality of the obtained scaling. As the scaling
is also well applicable to $\gtrsim$ 100 electrons, our results give interesting
prospects for the development of correlation functionals within density-functional 
theory.
\end{abstract}
\pacs{73.21.La, 78.67.Hc}
\maketitle

\section{Introduction\label{sec1}}

Artificial atoms -- quantum dots (QD) \cite{Hawrylak, Takag}, i.e., nanoscopic semiconductor 
structures where a set of electrons is confined, offer wider possibilities of engineering their 
properties than real atoms. The QD size, for example, can be changed from a few nanometers to 
hundreds of nanometers by modifying the experimental constraints. In turn, the degree of correlation 
of the electronic motion can be tuned by changing the system size, number of electrons, and
the type of the confining potential.

In the present paper, we focus on the standard, quasi-two-dimensional (quasi-2D), isotropic harmonic oscillator 
model of an artificial atom, given by the $N$-electron Hamiltonian
\begin{equation}
H=\sum^N_{i=1}\left[-\frac{\hbar^2}{2m^*}\nabla^2_i+V_{\rm ext}({r}_i)\right]
+\sum^N_{i<j}\frac{e^2}{\epsilon|{\mathbf r}_i-{\mathbf r}_j|} \ ,
\label{ham}
\end{equation}
where the external confinement in the {\it xy} plane is described by a parabolic potential,
\begin{equation}
V_{\rm ext}(r)=\frac 12 m\omega^2 r^2 \ .
\end{equation}
In spite of its simplicity, the model has been shown to predict 
very well the electronic properties of both vertical and lateral single quantum dots at the
GaAs/AlGaAs interface~\cite{Hawrylak,impurity,spindroplet}. The model can be essentially characterised
by two parameters, the number of confined electrons $N$ and the ratio between
Coulomb and oscillator energies, i.e.,
\begin{equation}
\beta = \frac{E_{\mathrm{Coul}}}{\hbar\omega}= \frac{e^{2}m^{1/2}}{4\pi\varepsilon\omega^{1/2}\hbar^{3/2}},
\end{equation}
where $e$ is the electron charge, $m$ its effective mass in the semiconductor material, and $\varepsilon$ 
the dielectric constant. We point out that $E_{\mathrm{Coul}}$ is the total electron-electron 
interaction energy, i.e., the expectation value of the Coulomb interaction operator. 

In a previous work \cite{universality}, some of the present authors showed that the total energy of 
such an artificial atom with $20\le N\le 90$ electrons obeys the scaling relation 
dictated by Thomas-Fermi (TF) theory
\begin{equation}
\frac{E_{\rm gs}(N,\beta)}{\hbar\omega}\approx N^{3/2} f_{gs}(z).
\label{egs}
\end{equation}
The variable $z=N^{1/4}\beta$ combines in a particular way the number
of electrons $N$ and the coupling constant $\beta$.  The function
$f_{gs}$ is {\em universal} in the sense that it depends only on $z$,
and not explicitly on the system parameters.  This result is not very
surprising, as the TF theory is known to predict well the total
energies of ``large'' electronic systems. In particular, in 2D the
gradient corrections to the TF kinetic energy vanish to all
orders~\cite{tf1}, and, secondly, for the 2D Fermi gas in a harmonic
potential the TF (non-interacting) kinetic energy is exact when using
the exact density as the input~\cite{tf2}.  It is amazing, however,
that the TF scaling for the total energy holds for a wide range of
$\beta$, from the strong-confinement (weak correlations,
$\beta\rightarrow 0$) to the weak-confinement limit (strong
correlations, $\beta\rightarrow\infty$, the so-called Wigner
phase~\cite{vignale}).

In the present paper, we move a step further and examine whether a scaling {\it a la} Thomas-Fermi, 
with different exponents, holds also for the correlation energy \cite{lowdin1,lowdin2,szabo} $E_c$, 
i.e., the difference between the total ground-state energy $E_{\rm gs}$ and the Hartree-Fock (HF) energy $E_{\rm HF}$:
\begin{equation}
E_c = E_{\rm gs} - E_{\rm HF}.
\label{ecor_def} 
\end{equation}
We point out that within density-functional theory (DFT) the HF energy in the above expression is commonly 
replaced by the total energy of an exact-exchange calculation. In practice, this is very close to the HF energy. 

We perform extensive calculations for $E_{\rm gs}$ of
QDs with $6\leq N\leq 90$ and find a unique scaling relation for $E_c$. The
fact that $E_c$ scales is completely unexpected because, by definition, the electronic correlation 
is beyond mean-field properties \cite{lowdin1,lowdin2,ziesche1,ziesche2,ziesche3,gersdorf,ziesche4,guevara,shi,sagar,kais,juhasz}.
In addition, the asymptotic behaviour of the correlation {\em potential} accessible in DFT
has a highly nontrivial scaling~\cite{stefano}. As our second main result, 
we find that also the fraction of the total energy associated with $E_c$ scales in a universal way.

The paper is organised as follows. In the next section we briefly summarize the computational 
methods employed. In Sec. \ref{results}, we present and discuss results from numerical 
calculations of $E_c$. Finally, concluding remarks are given in Sec. \ref{sect_concluding}.

\section{Numerical Results\label{sect2}}

In order to validate our numerical values for $E_c$, we perform
extensive calculations by means of standard many-electron methods
including HF, DFT in the local-density approximation
(LDA)~\cite{Attaccalite02,Makkonen12}, variational quantum Monte Carlo
(VMC)~\cite{Foulkes01,Jastrow55,Harju97,Harju05,ari_wigner_prb}, and
full configuration interaction (FCI) method~\cite{Rontani}.  We have
used our own numerical codes with all the methods, but the shown HF
and some LDA results have been calculated with the OCTOPUS
code~\cite{octopus}.

For simplicity, we give the confinement strength $\omega$ and energies
in atomic units. These can be used to obtain the material-dependent
effective atomic units, i.e. the effective Hartree energies and and
effective Bohr radii correspond to
$E_{h}^\ast=(m^\ast/m_0)/(\varepsilon/\varepsilon_0)^2E_h$ and
$a_0^\ast=(\varepsilon/\varepsilon_0)/(m^\ast/m_0) a_0$, respectively.
Our choices for $\omega$ used below represent typical experimental
setups for GaAs quantum dots. For the GaAs case, the effective atomic
units correspond to $E_{h}^\ast\approx 11.1277\,\mathrm{meV}$ and
$a_0^\ast\approx 10.11\,\mathrm{nm}$

\begin{figure}
\begin{center}
\includegraphics[width=1.0\linewidth,angle=0]{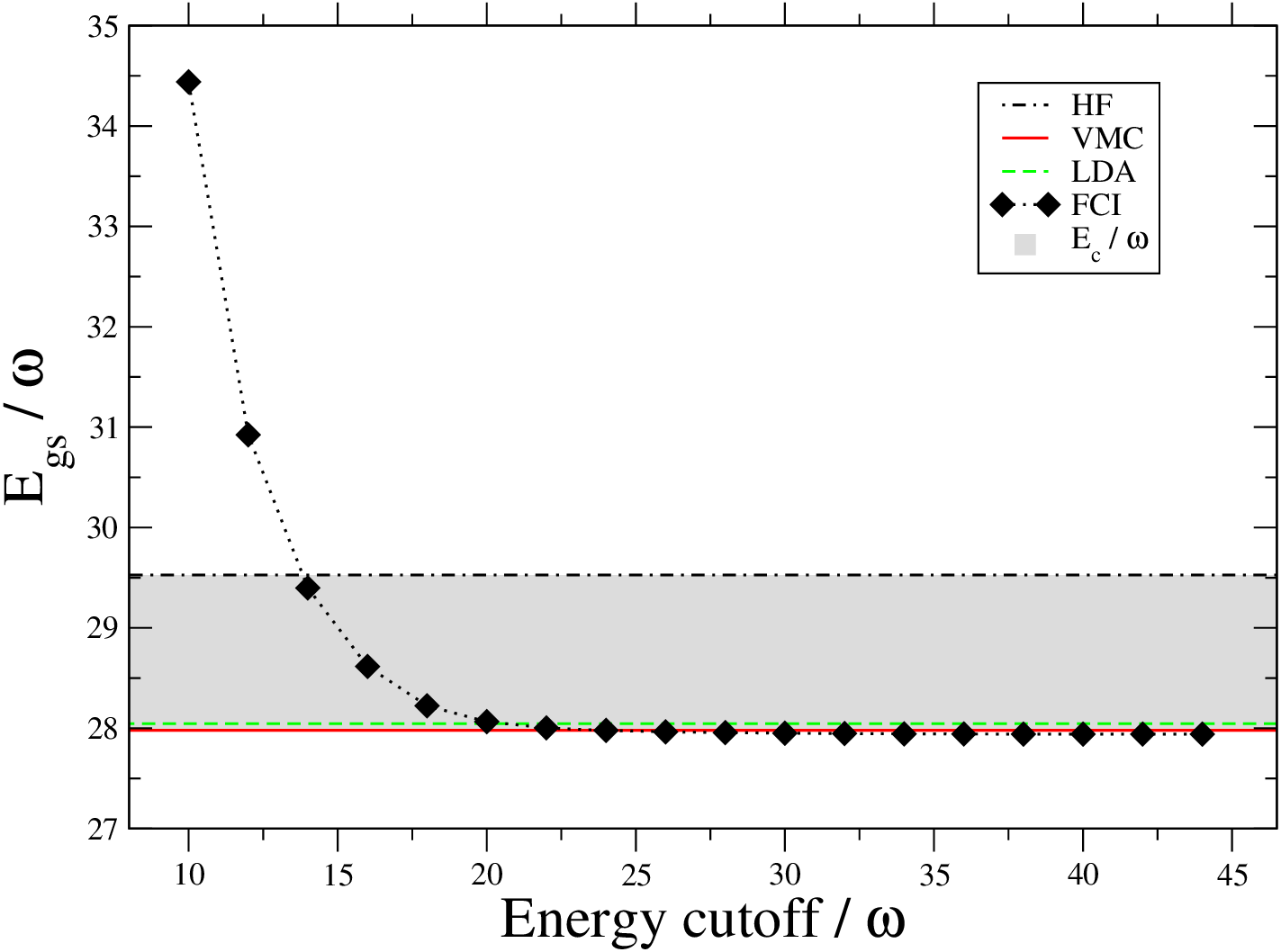}
\caption{\label{fig1} Convergence of FCI energies for a quantum dot
with $N=6$ and $\omega=0.25$ a.u. as a function of the non-interacting 
energy cutoff for the many-body configuration included to the basis \cite{Mikhailov02} (lines to guide the eye). 
When the basis is sufficiently large the VMC and LDA results coincide with the FCI result.
The difference to the HF result (higher in energy) corresponds to the correlation energy
(gray shaded region).}
\end{center}
\end{figure}

We find that both our VMC results (which are upper bounds for the total energy)
as well as the DFT results compare very well with FCI. This is demonstrated
in Fig.~\ref{fig1} for the case $N=6$ and $\omega=0.25$ a.u. that 
corresponds to the most correlated case of our set of QDs (see Fig.~\ref{fig2} below).
The FCI result is plotted as a function of the energy cutoff for the many-body
configuration included in the basis~\cite{Mikhailov02}, and it eventually converges to a value
that is very close to the VMC and LDA. It is noteworthy that the HF ground-state
energy (lacking the correlation) is higher, and thus the gray region
in Fig.~\ref{fig1} corresponds to the correlation energy according to 
Eq.~(\ref{ecor_def}).

The FCI method becomes too expensive with the larger particle numbers
considered. However, we can see from Fig. 1 that VMC is able to
capture nearly all of the correlation energy for the case
shown. Furthermore, this is the most correlated case we study in this
manuscript. We stress that the correlation energy is only a few
percent ($<$ 6 \%) of the total energy, reaching the largest relative
values for the smallest systems in the strong-coupling regime. This is
visualized in Fig.~\ref{fig2} that shows the relative correlation
energy $\chi = \vert E_c / E_{\rm gs}\vert$ as a function of the
confinement strength. The gray area corresponds to the typical
experimental regime when considering laterally or vertically confined
GaAs QDs.  One can see that the correlation energy is largest at small
particle number and with the weak confinements. The gray area in the
figure corresponds to the typical experimental regime when considering
laterally or vertically confined GaAs QDs.

\begin{figure}
\begin{center}
\includegraphics*[width=1.0\linewidth,angle=0]{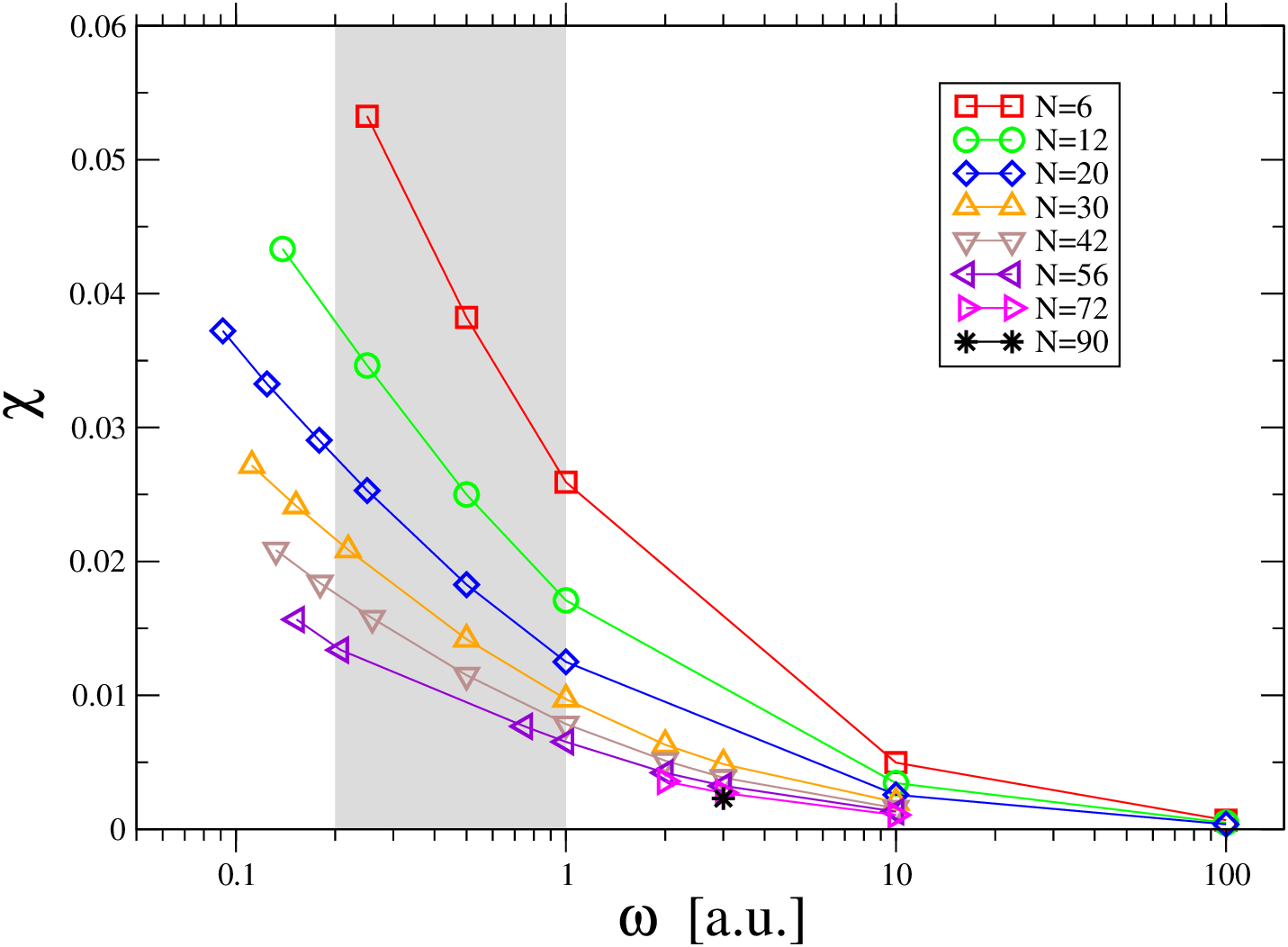}
\caption{\label{fig2} Relative correlation energies as a 
function of the confinement energy $\omega$ (log-scale). 
Symbols are the results from the VMC and LDA calculations. Lines are to guide the eye.
The gray shaded region corresponds to typical experimental setups for GaAs
quantum dots.}
\end{center}
\end{figure}

\begin{table*}
\caption{\label{table1}Hartree-Fock, LDA, and VMC ground-state energies of the quantum dots considered in this work.} 
\begin{indented}
\item[]\begin{tabular}{@{}lllll}
\br
$N$	&\hspace{1.2cm}$\omega$ [a.u.]	&\hspace{0.8cm}$E_{\rm HF}$ [a.u.]	&\hspace{0.8cm}$E_{\rm LDA}$ [a.u.]	&\hspace{0.8cm}$E_{\rm VMC}$ [a.u]	\\
\mr
\multirow{5}{*}{6} &\hspace{1.2cm}	0.25	&\hspace{0.8cm}	7.38845	&\hspace{0.8cm}	7.0114	&\hspace{0.8cm}	6.99496(8)\\
&\hspace{1.2cm}	0.5	&\hspace{0.8cm}	12.2713	&\hspace{0.8cm}	11.838	&\hspace{0.8cm}	11.8022(1)\\
&\hspace{1.2cm}	1	&\hspace{0.8cm}	20.7192	&\hspace{0.8cm}	20.252	&\hspace{0.8cm}	20.1821(2)\\
&\hspace{1.2cm}	10	&\hspace{0.8cm}	136.853	&\hspace{0.8cm}	136.61	&\hspace{0.8cm}	136.172(3)\\
&\hspace{1.2cm}	100	&\hspace{0.8cm}	1120.32	&\hspace{0.8cm}	1121.4	&\hspace{0.8cm}	1119.55(0)\\
\hline
\multirow{6}{*}{12} &\hspace{1.2cm}	0.138564	&\hspace{0.8cm}	16.1967	&\hspace{0.8cm}	15.485	&\hspace{0.8cm}	15.4946(1)\\
&\hspace{1.2cm}	0.25	&\hspace{0.8cm}	24.5034	&\hspace{0.8cm}	23.648	&\hspace{0.8cm}	23.6548(5)\\
&\hspace{1.2cm}	0.5	&\hspace{0.8cm}	40.2161	&\hspace{0.8cm}	39.217	&\hspace{0.8cm}	39.2110(9)\\
&\hspace{1.2cm}	1	&\hspace{0.8cm}	66.9113	&\hspace{0.8cm}	65.805	&\hspace{0.8cm}	65.7680(12)\\
&\hspace{1.2cm}	10	&\hspace{0.8cm}	416.192	&\hspace{0.8cm}	415.27	&\hspace{0.8cm}	414.759(8)\\
&\hspace{1.2cm}	100	&\hspace{0.8cm}	3248.39	&\hspace{0.8cm}	3249.3	&\hspace{0.8cm}	3246.84(3)\\
\hline
\multirow{8}{*}{20} &\hspace{1.2cm}	0.091268	&\hspace{0.8cm}	29.2580	&\hspace{0.8cm}	28.135	&\hspace{0.8cm}	28.1690(4)\\
&\hspace{1.2cm}	0.124230	&\hspace{0.8cm}	36.1435	&\hspace{0.8cm}	34.902	&\hspace{0.8cm}	34.9413(4)\\
&\hspace{1.2cm}	0.178885	&\hspace{0.8cm}	46.4969	&\hspace{0.8cm}	45.102	&\hspace{0.8cm}	45.1461(6)\\
&\hspace{1.2cm}	0.25	&\hspace{0.8cm}	58.6937		&\hspace{0.8cm}	57.157	&\hspace{0.8cm}	57.2088(11)\\
&\hspace{1.2cm}	0.5	&\hspace{0.8cm}	95.7327		&\hspace{0.8cm}	93.927	&\hspace{0.8cm}	93.9838(13)\\
&\hspace{1.2cm}	1	&\hspace{0.8cm}	158.004		&\hspace{0.8cm}	155.98	&\hspace{0.8cm}	156.030(1)\\
&\hspace{1.2cm}	10	&\hspace{0.8cm}	947.406		&\hspace{0.8cm}	945.44	&\hspace{0.8cm}	944.969(14)\\
&\hspace{1.2cm}	100	&\hspace{0.8cm}	7151.40		&\hspace{0.8cm}	7151.7	&\hspace{0.8cm}	7148.77(7)\\
\hline
\multirow{8}{*}{30} &\hspace{1.2cm}	0.111780	&\hspace{0.8cm}	67.1794	&\hspace{0.8cm}	65.274	&\hspace{0.8cm}	65.3545(9)\\
&\hspace{1.2cm}	0.152145	&\hspace{0.8cm}	82.9708	&\hspace{0.8cm}	80.876	&\hspace{0.8cm}	80.9676(12)\\
&\hspace{1.2cm}	0.219089	&\hspace{0.8cm}	106.689	&\hspace{0.8cm}	104.35	&\hspace{0.8cm}	104.463(1)\\
&\hspace{1.2cm}	0.5	&\hspace{0.8cm}	189.938	&\hspace{0.8cm}	187.09	&\hspace{0.8cm}	187.2425(8)\\
&\hspace{1.2cm}	1	&\hspace{0.8cm}	311.860	&\hspace{0.8cm}	308.65	&\hspace{0.8cm}	308.832(2)\\
&\hspace{1.2cm}	2	&\hspace{0.8cm}	519.252	&\hspace{0.8cm}	515.80	&\hspace{0.8cm}	515.976(2)\\
&\hspace{1.2cm}	3	&\hspace{0.8cm}	705.213	&\hspace{0.8cm}	701.68	&\hspace{0.8cm}	701.782(20)\\
&\hspace{1.2cm}	10	&\hspace{0.8cm}	1822.68	&\hspace{0.8cm}	1819.3	&\hspace{0.8cm}	1819.01(3)\\
\hline
\multirow{8}{*}{42} &\hspace{1.2cm}	0.132260	&\hspace{0.8cm}	133.470	&\hspace{0.8cm}	130.54	&\hspace{0.8cm}	130.687(2)\\
&\hspace{1.2cm}	0.180021	&\hspace{0.8cm}	164.821	&\hspace{0.8cm}	161.61	&\hspace{0.8cm}	161.793(2)\\
&\hspace{1.2cm}	0.259230	&\hspace{0.8cm}	211.875	&\hspace{0.8cm}	208.32	&\hspace{0.8cm}	208.536(3)\\
&\hspace{1.2cm}	0.5		&\hspace{0.8cm}	334.802	&\hspace{0.8cm}	330.78	&\hspace{0.8cm}	330.952(4)\\
&\hspace{1.2cm}	1		&\hspace{0.8cm}	547.683	&\hspace{0.8cm}	543.03	&\hspace{0.8cm}	543.381(10)\\
&\hspace{1.2cm}	2		&\hspace{0.8cm}	907.564	&\hspace{0.8cm}	902.50	&\hspace{0.8cm}	902.923(15)\\
&\hspace{1.2cm}	3		&\hspace{0.8cm}	1228.57	&\hspace{0.8cm}	1223.4	&\hspace{0.8cm}	1223.84(2)\\
&\hspace{1.2cm}	10		&\hspace{0.8cm}	3139.90	&\hspace{0.8cm}	3134.7	&\hspace{0.8cm}	3134.96(4)\\
\hline
\multirow{8}{*}{56} &\hspace{1.2cm}	0.152721	&\hspace{0.8cm}	239.599	&\hspace{0.8cm}	235.37	&\hspace{0.8cm}	235.846(5)\\
&\hspace{1.2cm}	0.2078699	&\hspace{0.8cm}	295.850	&\hspace{0.8cm}	291.24	&\hspace{0.8cm}	291.892(6)\\
&\hspace{1.2cm}	0.75		&\hspace{0.8cm}	722.112	&\hspace{0.8cm}	716.70	&\hspace{0.8cm}	716.563(14)\\
&\hspace{1.2cm}	1		&\hspace{0.8cm}	885.850	&\hspace{0.8cm}	879.88	&\hspace{0.8cm}	880.073(16)\\
&\hspace{1.2cm}	2		&\hspace{0.8cm}	1462.56	&\hspace{0.8cm}	1455.7	&\hspace{0.8cm}	1456.39(2)\\
&\hspace{1.2cm}	3		&\hspace{0.8cm}	1974.82	&\hspace{0.8cm}	1967.7	&\hspace{0.8cm}	1968.42(2)\\
&\hspace{1.2cm}	10		&\hspace{0.8cm}	5001.92	&\hspace{0.8cm}	4994.6	&\hspace{0.8cm}	4995.34(7)\\
\hline
\multirow{3}{*}{72} &\hspace{1.2cm}	2	&\hspace{0.8cm}	2218.71	&\hspace{0.8cm}	2210.8	&\hspace{0.8cm}	2210.78(3)\\
&\hspace{1.2cm}	3	&\hspace{0.8cm}	2989.61	&\hspace{0.8cm}	2980.8	&\hspace{0.8cm}	2981.51(4)\\
&\hspace{1.2cm}	10	&\hspace{0.8cm}	7516.12	&\hspace{0.8cm}	7506.3	&\hspace{0.8cm}	7508.14(12)\\
\hline
\multirow{1}{*}{90} &\hspace{1.2cm}	3	&\hspace{0.8cm}	4320.5	&\hspace{0.8cm}	4308.5	&\hspace{0.8cm}	4310.5(1)\\
\hline\hline
\end{tabular}
\end{indented}
\end{table*}

Our HF, LDA, and VMC results are summarized in Table~\ref{table1}. 
All the systems considered are closed-shell QDs with ground-state 
angular momentum and spin quantum numbers $L=S=0$.
As the VMC calculations are expected to yield the exact ground-state
energy with a good accuracy, we use those results in the following 
to compute the correlation energy as the difference from the HF energies. 
However, it is reassuring to see that the LDA energies are very close to
the VMC results. The accuracy of LDA in these systems is well expected
due to the smooth potentials and densities, and it is in accordance
with previous studies~\cite{spindroplet}.
We notice that -- for the parameters used in the calculations -- the scaled variable $z=N^{1/4}\beta$ 
takes values in the range $0<z<8$ that covers the weak-coupling and a part of the strong-coupling
regimes; $z\approx 1$ is the transition point between these regimes~\cite{GPP}. 

\section{Scaling of the correlation energy\label{results}}

Here, we first assume that the correlation energy has a particular scaling with respect to $N$. 
Thus, we suggest an ansatz of the form 
\begin{equation}
\frac{E_c}{\omega}=N^{\sigma}\,f_c(N^{1/4}\beta).
\end{equation}
Next, we fit our data for $N$ and $\beta$ to the ansatz by varying the scaling exponent $\sigma$.
We minimized the normalized root mean square deviation of
the data as a function of $\sigma$. The minimum deviation is obtained with $\sigma\approx 3/4$,
and thus we set $\sigma$ to this value. For the function $f_c$, a 
two-parameter fit of the form $\alpha z^{\gamma}$ leads to
\begin{equation}
\frac{E_c}{\omega\,N^{3/4}} = -0.0668\,z^{1.51}.
\label{ec_scaled} 
\end{equation}

\begin{figure}
\begin{center}
\includegraphics[width=1\linewidth,angle=0]{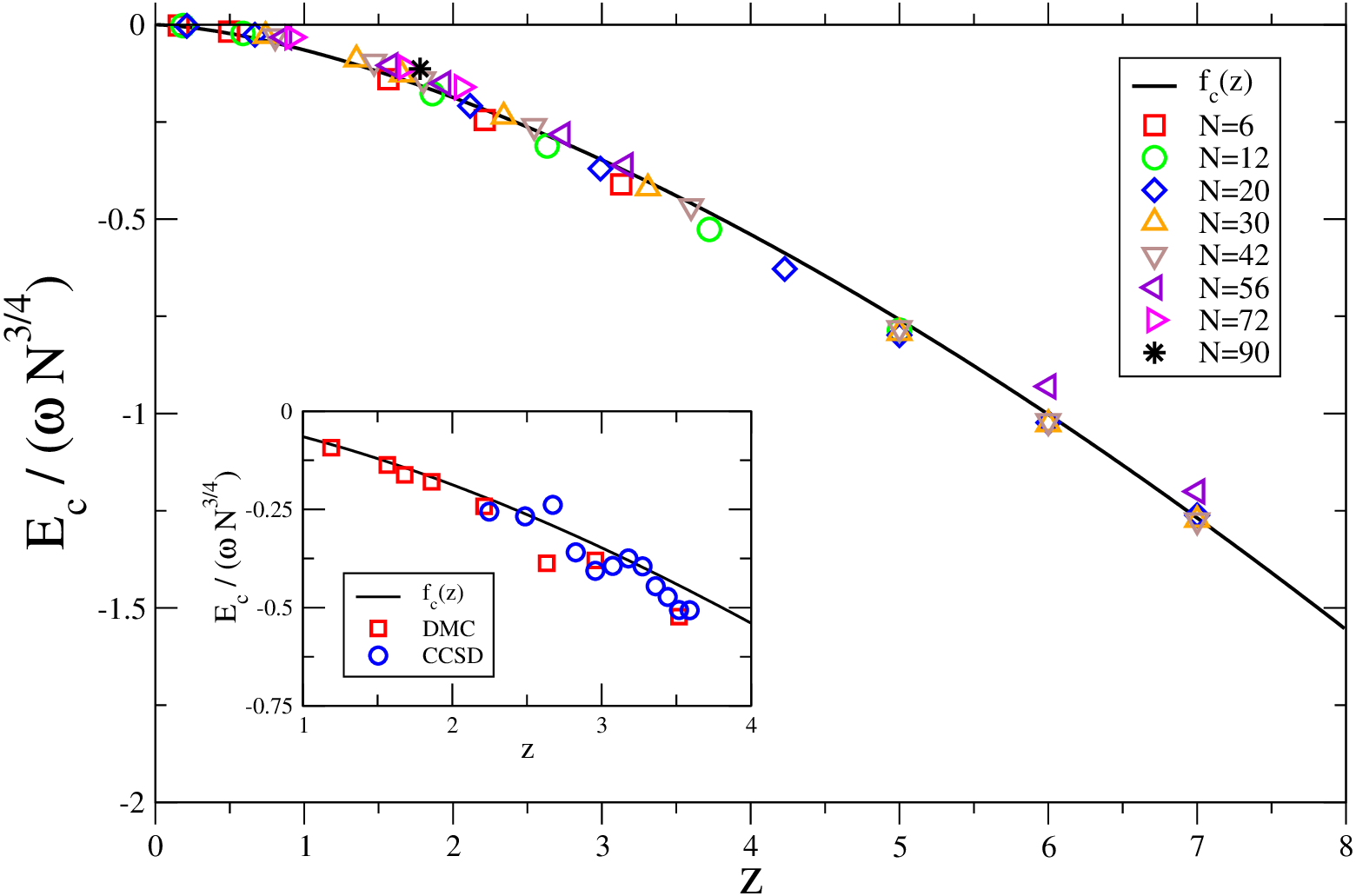}
\caption{\label{fig3} Scaled correlation energies as a function of 
the variable $z=\beta N^{1/4}$. Symbols correspond to VMC results and the solid curve 
represents the function $f_c$ in 
Eq.~(\ref{fig3}). Inset: results from diffusion Monte Carlo (DMC)~\cite{pederiva_2003}
calculations for $N=2\ldots 13$ and coupled-cluster singles-doubles (CCSD) calculations~\cite{waltersson} for $N=2,6,12$ 
compared against the scaling function $f_c$.}
\end{center}
\end{figure}

The scaled correlation energies are shown in Fig.~\ref{fig3} along with the 
function $f_c(z)$. It can be seen that the quality of the obtained analytical expression 
for $f_c$ is very good in the whole range of $z$, showing a mean deviation of 
about $5\%$ for $E_c$. We point out that Eq.~(\ref{ec_scaled}) can be straightforwardly 
rewritten in such a way that the dependence of $E_c$ on the system parameters 
(say, $N$ and $\omega$) becomes explicit:
\begin{equation}
 E_c(N,\omega) = -0.0668\ N^{1.1275} \omega^{0.245},
 \label{final}
\end{equation}
where $E_c$ is given in a.u. From this expression, we can trivially obtain an expression also for
the correlation energy per particle $E_c/N$.

Next we confirm the applicability of the obtained scaling relation by comparing
the scaling function $f_c$ in Eq.~(\ref{ec_scaled}) against independent
calculations for small and medium-size QDs, including open-shell systems. 
In particular, we consider diffusion Monte Carlo results for $N=2\ldots 13$ by 
Pederiva {\em et al.} \cite{pederiva_2003} and very recent coupled-cluster
 calculations for $N=2,6,12$ by Waltersson {\em et al.} \cite{waltersson}. The inset 
of Fig.~\ref{fig3} shows that both sets of results fit well with our scaling relation.
The largest deviations are found with smallest electron numbers ($N=2$).

Another interesting quantity to consider for the scaling is $\chi$ as defined above, i.e., 
the relative fraction of the correlation energy with respect to the total energy. 
It can be shown that $\chi$ also follows a similar scaling relation as a function of the parameter 
$z$. We use Eq.~(\ref{ec_scaled}) for $E_c$ and for the total ground-state energy we
use the result
\begin{equation}
\frac{E_{\rm gs}}{\hbar\omega N^{3/2}} = \frac{2}{3} + 
\frac{0.698\ z + 1.5\ z^{4/3} + 2.175\ z^{5/3}}{1 + 2.149\ z^{1/3} + 1.5\ z^{2/3} + 2.175\ z},
 \label{pade}
\end{equation}
obtained in Ref. \cite{universality}.
It has been found that the last expression performs better for systems with large $N$, 
in particular for $N>20$. For the fraction $\chi$, we obtain 
\begin{equation}
\chi(z) N^{3/4} = f_{\chi}(z) = \frac{p(z)}{q(z)},
 \label{rel_ecor_final}
\end{equation}
where
\begin{eqnarray}
 p(z) &=& 0.200 z^{1.513} + 0.431 z^{1.846} + 0.301 z^{2.180} \nonumber\\ 
      &\ & + 0.436 z^{2.513}.
\label{p}
\end{eqnarray}
and
\begin{eqnarray}
q(z) &=& 2 + 4.298 z^{1/3} + 3 z^{2/3} + 6.444 z \nonumber\\
     &\ & + 4.5 z^{4/3} + 6.525 z^{5/3}.
\label{q}
\end{eqnarray}

\begin{figure}
\begin{center}
\includegraphics[width=1.0\linewidth,angle=0]{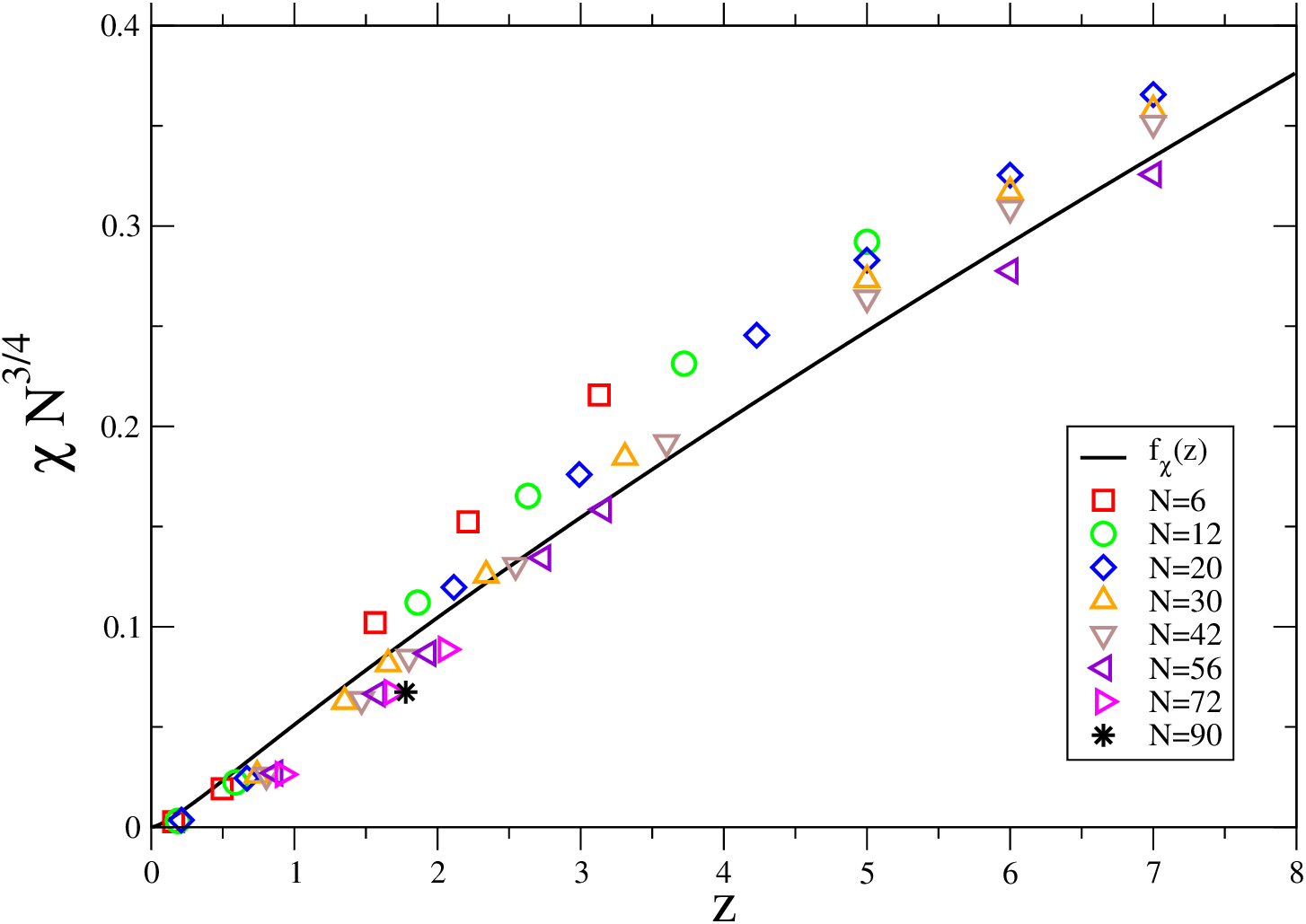}
\caption{\label{fig6} Scaled relative correlation energies 
$\chi = \vert E_c / E_{\rm gs}\vert$ as a function of the variable $z=\beta N^{1/4}$ obtained from the VMC
and LDA results. The solid line corresponds to Eq.~(\ref{rel_ecor_final}).}
\end{center}
\end{figure}

The computed values of $\chi$, labeled according to $N$, 
are shown in Fig. \ref{fig6} together
with the obtained analytic expression. It can be seen that Eq.~(\ref{rel_ecor_final}) 
works remarkably well for systems with $N>20$. Let us stress that in the weak-confinement 
regime ($z\gg 1$) Eq.~(\ref{rel_ecor_final}) predicts an almost linear dependence 
of $\chi N^{3/4}$ on the parameter $z$, leading to $\chi\sim N^{-0.54}\omega^{-0.42}$.

In the extreme low-density limit, the system approaches the so-called
Wigner phase~\cite{vignale}). The mean-field methods break the
rotational symmetry in this case, and the electron localize in space
\cite{Henkka}.  In the many-body treatment, the localization can be
seen in the conditional densities \cite{ari_wigner_prb}. Here, our
results are not in that limit, and analysis of this limit is left for future
studies.

Finally, we notice that in real (three-dimensional) atoms, 
the TF theory predicts for the total 
energy the following dependence~\cite{nuestroPRA}:
\begin{equation}
E_{\rm gs}(N,Z)\approx N^{7/3} f_{gs}(N/Z),
\label{egs_atoms}
\end{equation}
where $Z$ is the nuclear charge. The correlation energy also seems 
to show a scaling {\it a la} TF with
\begin{equation}
E_c(N,Z)\approx N^{\alpha} f_{c}(N/Z).
\label{ecor_atoms}
\end{equation}
The coefficient $\alpha$ is near 4/3 \cite{atoms,clementi1,clementi2,zolot,March92,nagy,mohajeri,liu,clementi3,kais-large-z}. 
We stress, however, that real atoms, unlike artificial ones, are always in the weak-correlation regime. 

\section{Concluding Remarks\label{sect_concluding}}

In summary, we have performed extensive numerical calculations for semiconductor quantum dots 
and found an unexpected universal scaling relation for the correlation energy, 
Eq.~(\ref{ec_scaled}), which resembles the scaling coming from TF theories. 
A universal scaling relation for the fraction of the total energy associated to the correlations was also obtained 
[Eq.~(\ref{rel_ecor_final})]. Such an expression provides information on the degree of correlation of the system and 
the accuracy of the HF estimation, even without any calculations. The material parameters (effective mass, 
dielectric constant) are contained in the scaling variable $z$. Our result has direct implications 
in the design of new correlation functionals 
for DFT calculations \cite{McCarthy,orbitalfree}, and may also supplement the recently founded DFT for
strictly correlated electrons~\cite{paola} and related approaches.

\ack
This work has been supported by the Academy of Finland through its Centres of Excellence Program 
(Project No. 251748) and through Project No. 126205 (E.R.). A.O. acknowledges support from CIMO, 
Finland (Gr. TM-11-7776) and is grateful to the members of the QMP Group (Aalto University School of Science) for their hospitality.
A.O and A.G. acknowledge support from the Caribbean Network for Quantum Mechanics, Particles and Fields (OEA, ICTP). 
A.D. and E.R. acknowledge support by the European Community’s FP7 through the CRONOS project, Grant Agreement No. 280879.

\end{document}